\documentclass[letter]{article}

\usepackage[english]{babel}
\usepackage[utf8]{inputenc}
\usepackage[T1]{fontenc}

\usepackage[letterpaper,top=3cm,bottom=2cm,left=3cm,right=3cm,marginparwidth=1.75cm]{geometry}

\usepackage{amsmath}
\usepackage{graphicx}

\usepackage[colorlinks=true, allcolors=blue]{hyperref}
\usepackage{stackengine}
\usepackage{physics}
\usepackage{amssymb} 
\usepackage{mathrsfs}
\usepackage{amsfonts}
\usepackage{amssymb}
\usepackage[useregional]{datetime2}
\usepackage[left]{lineno}
\usepackage{longtable}
\usepackage{booktabs}
\usepackage[framemethod=TikZ]{mdframed}
\usepackage{tcolorbox}
\tcbuselibrary{breakable}
\usepackage{csquotes}
\usepackage{harpoon}
\usepackage{enumitem} 
\usepackage{scalerel}
\usepackage{blindtext}
\usepackage[affil-it]{authblk}

\makeatletter
\let\ams@underbrace=\underbrace
\def\underbrace#1_#2{%
  \setbox0=\hbox{$\displaystyle#1$}%
  \ams@underbrace{#1}_{\parbox[t]{\the\wd0}{\raggedright#2}}%
}
\makeatother

\usepackage[backend=biber, maxcitenames=2, uniquename=false, style=apa, sorting=nyt,doi=false,isbn=false,url=false, natbib=true]{biblatex}
\DeclareNameFormat{labelname:poss}{% Based on labelname from biblatex.def
  \nameparts{#1}% Not needed if using Biblatex 3.4
  \ifcase\value{uniquename}%
    \usebibmacro{name:family}{\namepartfamily}{\namepartgiven}{\namepartprefix}{\namepartsuffix}%
  \or
    \ifuseprefix
      {\usebibmacro{name:first-last}{\namepartfamily}{\namepartgiveni}{\namepartprefix}{\namepartsuffixi}}
      {\usebibmacro{name:first-last}{\namepartfamily}{\namepartgiveni}{\namepartprefixi}{\namepartsuffixi}}%
  \or
    \usebibmacro{name:first-last}{\namepartfamily}{\namepartgiven}{\namepartprefix}{\namepartsuffix}%
  \fi
  \usebibmacro{name:andothers}%
  \ifnumequal{\value{listcount}}{\value{liststop}}{'s}{}}
\DeclareFieldFormat{shorthand:poss}{%
  \ifnameundef{labelname}{#1's}{#1}}
\DeclareFieldFormat{citetitle:poss}{\mkbibemph{#1}'s}
\DeclareFieldFormat{label:poss}{#1's}
\newrobustcmd*{\posscitealias}{%
  \AtNextCite{%
    \DeclareNameAlias{labelname}{labelname:poss}%
    \DeclareFieldAlias{shorthand}{shorthand:poss}%
    \DeclareFieldAlias{citetitle}{citetitle:poss}%
    \DeclareFieldAlias{label}{label:poss}}}
\newrobustcmd*{\posscite}{%
  \posscitealias%
  \textcite}
\newrobustcmd*{\Posscite}{\bibsentence\posscite}
\newrobustcmd*{\posscites}{%
  \posscitealias%
  \textcites}
  
\newcommand{\Cov}[3]{\mathrm{Cov}_{#1} \negmedspace \left( #2, #3 \right)}

\renewcommand{\eqref}[1]{Eq.\ref{#1}}

\def\big#1{{\hbox{$\left#1\vbox to15.5\p@{}\right.\n@space$}}}

\title{Towards a heuristic understanding of the storage effect}
\date{\today}
\author[1,2,*]{Evan C. Johnson}
\author[1]{Alan Hastings}

\affil[1,]{Department of Environmental Science and Policy; University of California Davis; Davis, California 95616 USA}
\affil[2,]{Center for Population Biology; University of California Davis; Davis, California 95616 USA}
\affil[*]{Corresponding author: Evan Johnson, evcjohnson@ucdavis.edu}

\begin{document}
\maketitle
\clearpage

\section*{Abstract}

The storage effect is a general explanation for coexistence in a variable environment. The generality of the storage effect is both a strength --- it can be quantified in many systems --- and a challenge --- there is not a clear relationship between the abstract conditions for storage effect and species' life-history traits (e.g., dormancy, stage-structure, non-overlapping generations), thus precluding a simple ecological interpretation of the storage effect. Our goal here is to provide a clearer understanding of the conditions for the storage effect as a step towards a better general explanation for coexistence in a variable environment. Our approach focuses on dividing one of the key conditions for the storage effect, covariance between environment and competition, into two pieces, namely that there must be a causal relationship between environment and competition, and that the effects of the environment do not change too quickly. This finer-grained definition can explain a number of previous results, including 1) that the storage effect promotes annual plant coexistence when the germination rate fluctuates, but not when the seed yield fluctuates, 2) that the storage effect is more likely to be induced by resource competition than apparent competition, and 3) that the spatial storage effect is more probable than the temporal storage effect. Additionally, our expanded definition suggests two novel mechanisms by which the temporal storage effect can arise: transgenerational plasticity, and causal chains of environmental variables. These mechanisms produce coexistence via the storage effect without any need for stage structure or a temporally autocorrelated environment. 

Keywords: storage effect, spatial storage effect, coexistence mechanisms, temporal autocorrelation, stage-structure, causation

\tableofcontents

\newpage

\section{Introduction}
\label{Introduction}

The storage effect is a general explanation for how species can stably coexist by specializing on different environmental states; it can be thought of as the formalization of environmental niche partitioning. Unfortunately, the storage effect is difficult to understand in its entirety. The problem is that the storage effect is a general phenomenon that can look very different in different models, thus making it difficult to relate the storage effect to a small set of ecological constructs such as dormancy, stage-structure, and environmental autocorrelation. For instance, generalizing from the results of the lottery model (a seminal model in which the ecological storage effect was discovered; \cite{chesson1981environmentalST}), one may be tempted to claim that the storage effect occurs when species have a robust life-stage that can "wait it out" for a good year. However, this interpretation turns out to be imprecise, since neither stage-structure nor overlapping generations are required for the storage effect. Another general interpretation of the the storage effect is that it requires rare species to be \textit{buffered} from the double whammy of a bad environment and high competition. This too turns out to be imprecise (\cite{johnson2022storage}). 

Perhaps a general ecological interpretation of the storage effect is too ambitious. Instead, we can gain insight by studying the \textit{ingredient-list definition of the storage effect}: a list of abstract conditions that \textit{tend} to lead to a systematically positive storage effect, i.e., a storage effect uplifts most species in a community. Here, we attempt to make the storage effect more understandable by expounding a single ingredient: the covariance between environment and competition. This paper is not meant to be a review of the storage effect, as this has been done elsewhere (\cite{johnson2022storage}). 

The ingredient-list definition states that the storage effect depends on 

\begin{enumerate}
    \item Species-specific responses to the environment,
    
    \item a non-zero interaction effect of environment and competition on per capita growth rates (also known as \textit{non-additivity}), and 
    
    \item covariance between environment and competition (\textit{$EC$ covariance}).
\end{enumerate}

The function of the ingredient 1 is rather obvious: species-specific responses to the environment establishes the presence of niche differences, which are always necessary for coexistence. In the context of ecological coexistence, the term "niche differences" usually refers to differences in resource consumption (\cite{tilman1982resourceST}), the affinities of natural enemies (\cite{holt1977predationST}), or social/behavioral differences (\cite{chesson1991need}). What makes the storage effect unique is that coexistence is achieved through \textit{environmental} niche differences.

Ingredient 2, an interaction effect between environment and competition, is akin to an interaction effect in a multiple regression where the response variable is the per capita growth rate, and the predictor variables are the environment and competition parameters. Functionally, the interaction effect can be thought of as combining the environment and competition into a large number of \textit{effective regulating factors} (analogous to resources or natural enemies) that species can specialize on (\cite{johnson2022storage}). 

However, this is all very abstract. What causes an interaction effect in particular ecological systems? In the seminal models of coexistence theory (the lottery model and the annual plant model; \cite{Chesson1994}) a robust life-stage / overlapping generations is necessary for an interaction effect. In other models, an interaction effect results from population structure, whether it be dormancy (\cite{caceres1997temporal}; \cite{ellner1987alternate}), phenotypic variation (\cite{chesson2000mechanisms}), or spatial population structure (\cite{chesson2000general}). However, an interaction effect can arise in models without population structure or overlapping generations, purely due to a multiplicative form of the per capita growth rate function (\cite{li2016effects}; \cite{letten2018species}; \cite{Ellner2019}). It is also worth noting that the storage effect was originally discovered by population geneticists, and that in the population genetic version of the storage effect, an interaction effect can result from heterozygosity (\cite{Dempster1955}; \cite{Haldane1963}), sex-linked alleles (\cite{reinhold2000maintenance}), epistasis (\cite{gulisija2016phenotypic}), and maternal effects (\cite{yamamichi2017roles}). In summary, there are many ways for an interaction effect to occur. At least for the moment, it is not possible to give a general interpretation of the interaction effect in terms of life-history characteristics (e.g., dormancy, a robust life-stage, phenotypic variation).

The final ingredient, covariation between environment and competition, is the focus of this paper.  Because covariation is usually thought of as a statistical measure of linear association, it is not clear how it is likely to arise in real communities. To make ingredient 3 more comprehensible, we split it into two sub-ingredients: 3A) a causal relationship between environment and competition (i.e., a good environment leads to high competition, or conversely, a bad environment leads to low competition), and 3B) that the effects of the environment do not change too quickly, relative to the rate at which the environment affects competition. This finer-grained list can be levied to understand a number of theoretical results, and to intuit novel mechanisms through which the storage effect can arise.

\section{Expanding the ingredient-list definition of the storage effect}
\label{Expanding the ingredient-list definition of the storage effect}

\begin{figure} 
 \centering
      \includegraphics[scale = 0.5]{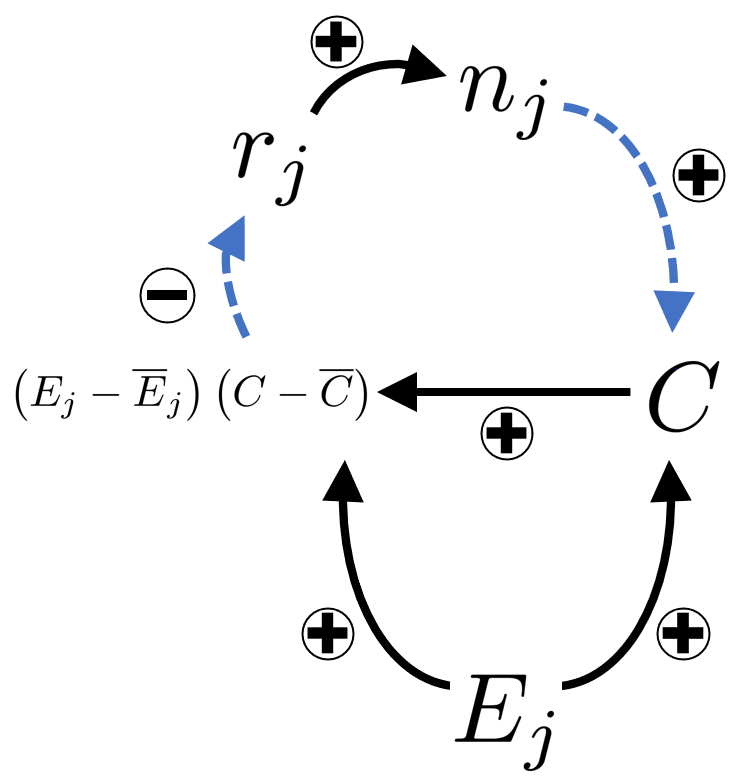}
  \caption{A causal diagram of the storage effect. $j$ is the species index, $r_j$ is the per capita growth rate, $n_j$ is population density, $E_j$ is the species-specific environmental parameter, $C$ is competition,  and $\left(E_j-\overline{E}_j\right)\left(C-\overline{C}\right)$ is an effective regulating factor that becomes $\Cov{}{E_j}{C_j}$ when averaged across time. The black arrows show the direction of causation, e.g., an increased per capita growth rate $r_j$ causes increased population density $n_j$ in the future. The blue dashed arrow shows a non-causal nested relationship. For example, $r_j$ is a function \textit{of} $\left(E_j-\overline{E}_j\right)\left(C-\overline{C}\right)$, with the negative sign showing that $r_j$ is decremented by this effective regulating factor due to the negative interaction effect. The species-specific response to the environment, $E_j$, serves two functions. 1) $E_j$ affects  $C$ --- a good environment causes to high competition --- and does not change too quickly, thus ensuring that the term $\left(E_j-\overline{E}_j\right)\left(C-\overline{C}\right)$ is non-zero when averaged across time. 2) $E_j$ is \textit{species-specific}, which effectively turns a single regulating factor, $C$, into a great number of regulating factors, $\left(E_j-\overline{E}_j\right)\left(C-\overline{C}\right)$ (potentially one for each species), thus allowing for sort of specialization that is necessary for coexistence.}
  \label{causal diagram}
\end{figure}

The ingredient-list definition of the storage effect can be expanded as follows:

\begin{enumerate}[label=3.]
\item Covariance between environment and competition.
\begin{enumerate}[label=3\Alph*.]
\item  A causal relationship between environment and competition, and
\item the effects of the environment do not change too quickly, relative to the rate at which the environment affects competition.
\end{enumerate}
\end{enumerate}

Before proceeding, we must note that the terms "environment" and "competition" are used loosely. The "environment" can represent an abiotic variable (e.g., temperature), or a demographic parameter that depends on abiotic variables (e.g., germination probability depends on temperature), or more generally, the effects of density-independent factors. Due to this generality, the environment has also been called the "environmental response" or the "environmentally-dependent parameter". Similarly, competition can be more generally understood as the effects of regulating factors, which may include species' densities, resources, refugia, territories, natural enemies, etc.

The purpose of the first sub-ingredient, 3A, is to show that the environment $E$ "goes along with" competition $C$, because $E$ (in part) causes $C$. Causation is necessary for correlation in this context (i.e., models of population dynamics) because there are no latent variables (also known as third variables) that could affect both $E$ and $C$, and therefore produce a spurious correlation. 

The purpose of the second sub-ingredient, 3B, is more difficult to understand. Per capita growth rates depend on the current values of $E$ and $C$, via the term $(E(t) - E^*)(C(t) - C^*)$ (where $E^*$ and $C^*$ are the equilibrium levels of these variables). However, since the environment causally affects the level of competition, and causes precede their effects, the only guaranteed statistical relationship is that between the current value of $C$ and the past value of $E$ (i.e., $(E(s) - E^*)(C(t) - C^*) > 0$, for some $s < t$). Figure \ref{causal diagram} illustrates this idea: one causal arrow (and thus one unit of time) is required for the environment to directly affect growth rates, whereas two causal arrows (and thus two units of time) are required for the effects of the environment on the growth rate to be mediated through competition. For a non-zero covariance between the current environment and competition, it is essential that the effects of the environment are carried forward through time, such that the effect of a past environment is brought into contact with the competition that it caused.

Ingredient 3B is perhaps the most surprising thing about the storage effect. It seems natural for species' responses to the environment to \textit{not} be perfectly correlated (satisfying ingredient 1). Even if species are subjected to strong convergent evolution or or environmental filtering, we would still expect some systematic difference between species due to evolutionary transient dynamics, development constraints, etc. It also seems natural for species to experience an interaction effect between environment and competition (satisfying ingredient 2), seeing as how the alternative --- additivity --- takes a very specific form in scalar populations: $\lambda = n(t+1)/(n(t) = \exp{\alpha E + \beta_j C + c}$, where $\alpha$, $\beta$, and $c$ are constants. \textcite{Chesson1994} writes "There are so many ways in which nonadditivity can arise that it seems doubtful that any real populations could be additive,...". Finally, it seems natural for a good environment to cause high competition (satisfying ingredient 3A) as initially high population growth leads to overcrowding (\cite{chesson1988community}). However, there is no  guarantee that the environment won't change more quickly than the time it takes for its causal effects on competition to be felt. To show this more explicitly, we analyze a toy model and find that the covariance between environment and competition is proportional to $T_E / T_{E \rightarrow C}$, where $T_E$ is the timescale of environmental autocorrelation, and $T_{E \rightarrow C}$ is the timescale at which the environment affects competition.

To keep things as simple as possible, we analyze a single-species model; this can be thought of as part of an invasion analysis for a two-species community. The time-evolution of population dynamics is given by the relation $n(t+dt) = n(t) + F(n(t),E(t)) \, dt$, where $F$ is a population growth rate function, $n$ is population density, $E$ is the environmental parameter, $t$ is time, and $dt$ is the length of a time-step. The time evolution of $E$ is given by the relation $E(t+dt) = E(t) + G(E(t)) \, dt + \sigma \, dW(t)$, where $G$ is the deterministic change function, $\sigma$ is the scale of environmental fluctuations, and $dW$ is an increment of the standard Wiener process (\cite{karlin1975first}). 

Suppose that in the absence of fluctuations in $E$ (i.e., in the limit as $\sigma \rightarrow 0$) the system would come to a stable equilibrium where the state is $n^*$ and $E^*$. Suppose further that $\sigma$ is very small (relative to other parameters hidden in $F$ and $G$). Then, we can use a \textit{small-noise approximation} (\cite{gardiner1985handbook}) to approximate the dynamics of $n$ and $E$ about the equilibrium point. The resulting equations are

\begin{equation}
\begin{aligned}
    d n & = \left[ \frac{\partial F(n^*,E^*)}{\partial n} (n-n^*) + \frac{\partial F(n^*,E^*)}{\partial E} (E-E^*)  \right] d t \\
    d E & = \frac{\partial G(E^*)}{\partial E}(E-E^*) d t + \sigma \, d W,
\end{aligned}
\end{equation}

where the partial derivatives are first calculated symbolically and then evaluated at the equilibrium point, as the notation implies. Despite $F$ and $G$ being arbitrary functions, the population dynamics take a simple form: the equation for the time-evolution of $n$ is a linear Langevin equation, and the equation for the time-evolution of $E$ is an Ornstein-Uhlenbeck process. The covariance between $E$ and $n$ can be calculated with the help of Ito's lemma and the Ito-Isometry Principle (\cite{karlin1981second}). For convenience, we use the program \textit{Mathematica} (see  {\fontfamily{qcr}\selectfont EC\_cov.nb} at \url{https://github.com/ejohnson6767/storage_effect_heuristic}. In the stationary joint stationary distribution, the covariation between environment and population density is 

\begin{equation}
\Cov{}{E}{n} = \frac{\frac{\partial F(n^*,E^*)}{\partial E} \sigma^2}{2 \left(\frac{\partial G(E^*)}{\partial E} \left(\frac{\partial G(E^*)}{\partial E} + \frac{\partial F(n^*,E^*)}{\partial n} \right)\right)}.
\end{equation}

Suppose that the competition parameter is a function $H$ of current population density, $C(t) = H(n(t))$, as is the case in the classic Lotka-Volterra model, the multi-species Ricker model (\cite{dallas2021initial}), the Hassel model (\cite{hassell1976discrete}), the Beverton-Holt competition model Ackleh  (\cite{walters1999linking}; \cite{ackleh2005discrete}), the annual plant model (\cite{chesson1990geometry}; \cite{Chesson1994}; \cite{lanuza2018opposing}), the lottery model (\cite{Chesson1994}; \cite{Yuan2015}), and other related models (\cite{brauer2012mathematical}).  Now, we can approximate fluctuations in the competition parameter as $(C-C^*) \approx \frac{\partial H(n^*)}{\partial n} (n-n^*)$, and thus, 

\begin{equation}
\Cov{}{E}{n} = \frac{\frac{\partial C(n^*)}{\partial n} \frac{\partial F(n^*,E^*)}{\partial E} \sigma^2}{2 \left( \frac{\partial G(E^*)}{\partial E} \left(\frac{\partial G(E^*)}{\partial E} + \frac{\partial F(n^*,E^*)}{\partial n}\right)\right)}.
\end{equation}

We will now re-parameterize the covariance in terms of characteristic time-scales. The rate at which the environmental response decays to equilibrium is $-\partial G(E^*) / \partial E$, so the characteristic timescale of environmental change is $T_E = - 1 / \frac{\partial G(E^*)}{\partial E}$. The rate at which fluctuations in $E$ positively affects $C$ is $ - \frac{\partial H(n^*)}{\partial n} \frac{\partial F(n^*,E^*)}{\partial E}$, so the characteristic timescale at which the environment affects competition is $T_{E \rightarrow C} = - 1 / \left( \frac{\partial H(n^*)}{\partial n} \frac{\partial F(n^*,E^*)}{\partial E} \right)$. 

The covariance can now be written as 

\begin{equation}
\Cov{}{E}{C} = \frac{\left(T_E \sigma \right)^2 }{2 \, T_{E \rightarrow C} \left(1 - T_E \, \frac{\partial F(n^*,E^*)}{\partial n}\right) }
\end{equation}

which succinctly shows that the covariance increases monotonically with the ratio $T_E / T_{E \rightarrow C}$ (note that $\frac{\partial F(n^*,E^*)}{\partial n}$ is negative, so the denominator is always positive). In words, a positive covariance between environment and competition requires that environmental correlations last longer than the time time it takes the environment to appreciably affect competition. 

\section{Discussion}

Ingredient 3B can explain a couple of interesting theoretical findings about the storage effect. \textcite{Kuang2009CoexistenceEffect} analyzed a model in which two species had one shared resource and one shared predator. Resource competition generated a storage effect, whereas the shared predator did not. Ingredient 3B explains why. The time-scale of environmental change is a single time-step, but the time it takes for the environment to affect predator density is two time-steps: one time-step for the environment to affect prey density, and one time-step for prey density to affect predator density. In contrast, a predator-mediated storage effect may arise if predators respond quickly to prey density, as is the case with prey-switching behavior (\cite{kuang2010interacting}; \cite{chesson2010storage}) or satiation due to a type 2 functional responses (\cite{stump2017optimally}). 

Another interesting result is that in the annual plant model, (\cite{Chesson1994}) the storage effect arises when germination probability fluctuates, but not when the seed yield fluctuates. Ingredient 3A --- a causal relationship between environment and competition --- is satisfied if either germination or yield fluctuates (i.e., is identified as the environmental parameter $E$). Increased per germinant yield $Y$ increases the density of seeds $X$, which increases the number of germinants $X G$, which increases the level of competition. Increased germination $G$ leads to increased germinants $X G$, which increases the level of competition. However, note the difference in the length of the two causal pathways: the germination probability affects competition in the current time-step, whereas the yield affects competition in the next time-step; by then, the environment has changed, such that ingredient 3B is not satisfied, and thus the covariance between environment and competition (a.k.a. $EC$ covariance) evaporates.

Ingredient 3B --- carrying the effects of the environment forward through time --- can be thought of a novel type of storage. The environment is "stored" in an autocorrelated environment (\cite{loreau1989coexistence}; \cite{loreau1992time}; \cite{li2016effects}; \cite{schreiber2021positively}), since current growth rates will be predictive of future growth rates. In the lottery model with only temporal variation, the effects of the environment are "stored" in larvae which disperse to the pelagic zone for weeks or months (\cite{green2015larval}). Note that in the lottery model, the classical notion of storage (i.e., "buffering" via a robust life-stage) is about generating an interaction effect (ingredient 2) via long-lived adult fish; the novel notion of storage (i.e., carrying the effects of the environment through time) is about generating a covariance (ingredient 3) through the comparatively short-lived larvae.  

To date, all models of the temporal storage effect feature either temporal autocorrelation or stage-structure, although the latter is sometimes implicit, as is the case in the lottery model and annual plant model (\cite{Chesson1994}). However, once one accepts that the primary purpose of these constructs is to satisfy ingredient 3B, it becomes readily apparent that the storage effect can arise in other situations. Here, we present two novel mechanisms that enable the storage effect, neither of which require temporal autocorrelation nor stage-structure. 

First, we contend that transgenerational plasticity (e.g., maternal effects, epigenetics) can carry the effects of the environment forward through time, therefore satisfying ingredient 3B.  Note that what we are proposing here is different from the the model of \cite{yamamichi2017roles}, where maternal effects (a type of transgenerational plasticity) produces a negative interaction effect and diploidy leads to the $EC$ covariance. Even though transgenerational plasticity can generate an $EC$ covariance, plasticity of any type is not likely to evolve in a quickly changing environment (\cite{stomp2008timescale}). Therefore, it may be interesting to use the adaptive dynamics framework (\cite{geritz1998evolutionarily}; \cite{brannstrom2013hitchhiker}) to study the evolution of the storage effect due to transgenerational plasticity.

\begin{figure} 
 \centering
      \includegraphics[scale = 0.5]{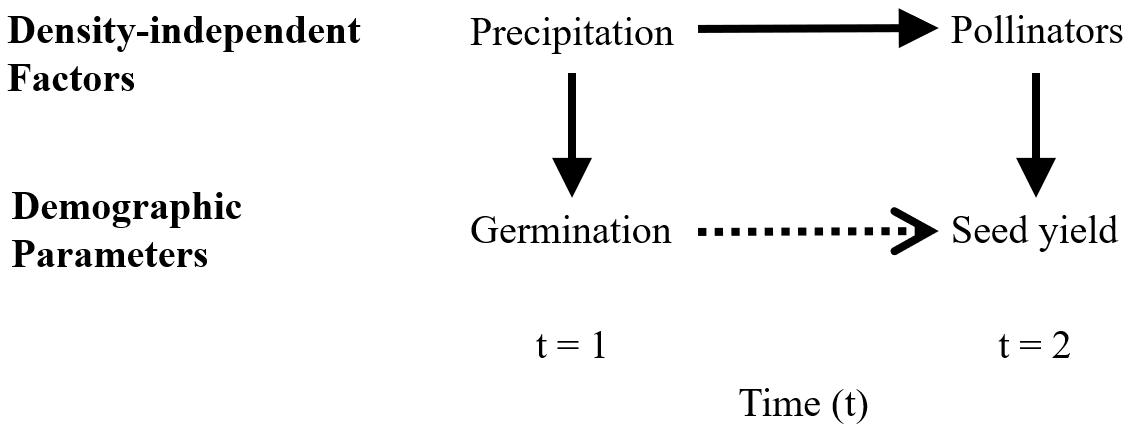}
  \caption{The covariance between environment and competition can be generated by causal chains of environmental variables. Solid arrows denote causal relationships. The dotted arrow denotes a non-causal, indirect relationship. The causal relationship between the exogenous density-independent factors --- precipitation and pollinators --- prevents the effects of the environment from changing too quickly, thus satisfying ingredient 3B. The demographic parameters are correlated because both are causally affected by precipitation on different time-lags.}
  \label{causal_chain}
\end{figure}

Second, we contend that causal chains of environmental responses can satisfy ingredient 3B (Fig \ref{causal_chain}). Consider a community of annual plants. High precipitation in year 1 causes a high germination probability in year 1, and thus a large number of germinants in year 2. Simultaneously, high precipitation in year 1 causes a high abundance of fly pollinators in year 2, which causes a high per germinant seed yield in year 2. Thus, there is a covariance between an environmental response (i.e., per germinant seed yield) and competition (i.e., the density of germinant competitors), even if the abiotic environment (precipitation) and species' environmental responses (germination probability and per germinant yield) are temporally uncorrelated. 

The previous example can be explained in two ways, depending on how one understands "the environment". In MCT, it is conventional for "the environment" to be a demographic parameter that depends on fluctuating density-independent factors. If we take this perspective, then it is clear that there is not a causal relationship between the environmental parameters, germination and yield. Rather, there is an indirect relationship that is a consequence of both parameters ultimately being caused by precipitation, but with different time-lags (Fig \ref{causal_chain}). If on the other hand, we identify "the environment" as exogenous density-independent factors, then the $EC$ covariance (more specifically, ingredient 3B) is generated by a causal chain of environmental variables, wherein precipitation causes increases in the pollinator population.

Ingredient 3B also explains the putative potency of the spatial storage effect, which "seems to be inevitable under realistic scenarios" (\cite{chesson2000general}). In models with permanent spatial heterogeneity, the local environment does not change over time, thus automatically satisfying ingredient 3B. This is not to say that environmental heterogeneity guarantees an environmental-competition covariance. It must also be the case that not all individuals disperse after every time-step. This \textit{local retention} allows populations to build up in good environments, thus satisfying ingredient 3A: a causal relationship between the local environment and local competition. It is interesting to note that the primary contingency for the temporal storage effect is ingredient 3B (will the effects of the environment be carried through time?) whereas the contingency for the spatial storage effect is 3A (is the spatial scale of patches smaller than the scale of dispersal, such that the local environment has a causal relationship with local competition?). 

The most thorough empirical test of the spatial storage effect found near-zero $EC$ covariances in a community of woodland annual forbs, grasses and geophytes (\cite{towers2020requirements}). The authors provide several reasons for the absence of covariance, but ingredient 3A suggests an additional reason. It is possible that the average dispersal distance of the plants (1-3 meters (\cite{harper1977population}), or much more with flooding; \cite{gutterman2000environmental}) is much greater than the grain size of environmental variation; in some systems, resource availability can vary significantly across a meter (\cite[p.~100]{tilman1982resourceST}; \cite{bogunovic2014spatial}). If this is the case, species will not be able build up populations in locations where the environment is favourable.

Even if there is no local retention, population buildup can occur when survival or mortality fluctuates across space. In the annual plant model with no local retention and global dispersal, A high yield in a particular patch does not lead to increased competition in that patch, because new seeds are distributed evenly over the landscape. However, a patch with a high seed survival probability will lead to a buildup of the local seed-bank, thus leading to increased local competition after seeds germinate. The sedentary seed-stage behaves like local retention, in the sense that both satisfy ingredient 3A. The same could be said of the non-dispersing adult fish in the lottery model. However, in both the lottery model and the annual plant model, there is no interaction effect (ingredient 2 is not satisfied) when the survival probability is identified as the spatially-fluctuating environmental response. Note: this is not true in the context of calculating the temporal storage effect, due to the fact that temporal coexistence mechanisms are calculated by decomposing the log-transformed finite rate of increase, $r = \log(\lambda)$, whereas spatial coexistence mechanisms are calculated by decomposing $\lambda$ (\cite[p.~218]{chesson2000general}). While spatial variation in survival does not engender a storage effect (at least in some simple models), the variation in population density that results from differential population buildup can engender \textit{fitness-density covariance} (see \cite{muko2000species} for an example), a related coexistence mechanism that is outside the scope of this paper.

The storage effect is one of the most important concepts in community ecology. It subverted the ecology milieu of the 1970s, which focused on coexistence via resource partitioning and regarded environmental stochasticity as a malignant force, both for individual species' persistence (\cite{Lewontin1969}) and for multi-species coexistence (\cite{May1974}). Further, the storage effect subverted a tradition of thought going back to Darwin, who viewed competitive exclusion as the status quo of nature (see \cite{lewens2010natural} for the reasons why), and therefore, that coexistence was the oddity worth explaining: "We need not marvel at extinction; if we must marvel, let it be at our own presumption in imagining for a moment that we understand the many complex contingencies on which the existence of each species depends." (\cite[p.~322]{darwin1859origins})

Darwin's presumption of competitive exclusion was formalized by the \textit{competitive exclusion principle} (\cite{volterra1926variationsST}, \cite{Lotka1932TheSupply}, \cite{GauzeG.F.GeorgiĭFrant︠s︡evich1934TsfeST}; \cite{levin1970community}), which stated that no more than $N$ species can coexist on $N$ resources, and later brought into focus by \posscite{hutchinson1961paradox} \textit{paradox of the plankton}, which asked how dozens of lake phytoplankton species could coexist on a handful of limiting nutrients. By showing that an arbitrary number of species can coexist on a single resource  (e.g., \cite[Eq.~81]{Chesson1994}), the storage effect flipped the question of "Why are there so many species?" to "Why is the number of species that which we observe?" To this end, the storage effect and other coexistence mechanisms have been measured in a number of real ecological communities (\cite{caceres1997temporal}; \cite{venable1993diversity}; \cite{pake1995coexistence}; \cite{pake1996seed}; \cite{adler2006climate}; \cite{sears2007new}; \cite{descamps2005stable}; \cite{facelli2005differences}; \cite{angert2009functional}; \cite{Adler2010}; \cite{usinowicz2012coexistence}; \cite{chesson2012storage}; \cite{chu2015large}; \cite{usinowicz2017temporal}; \cite{ignace2018role}; \cite{hallett2019rainfall}; \cite{armitage2019negative}; \cite{armitage2020coexistence}; \cite{zepeda2019fluctuation}; \cite{zepeda2019fluctuation}; \cite{holt2014variation}; \cite{ellner2016quantify}; \cite{Ellner2019}).

Surely, such a historically and currently important concept deserves to be understood. In this paper, we have attempted to provide a better heuristic explanation of the storage effect by showing how an $EC$ covariance is likely to arise. Our analysis shows how seemingly disparate models are actually similar. For example, a juvenile life-stage (e.g. larvae in the lottery model), environmental autocorrelation, and spatial heterogeneity all serve the same function: carrying the effects of the environment forward through time, to bring it into contact with the competition that it caused.

Future research should focus on further explicating ingredient 2, an interaction effect between environment and competition. The interaction arises from a variety of mechanisms in a variety of models (see the \nameref{Introduction}), and it is unclear what ties these mechanisms together. For example, \textcite{schreiber2021positively} used a very simple (and thus ostensibly general) model in which fluctuating survival drives a positive interaction effect, but fluctuating fecundity drives a negative interaction effect. The storage effect would be much more understandable and predictable if one could know the sign of an interaction effect based only on a verbal description of an ecological system, not a mathematical analysis or detailed background knowledge about different classes of models. 

\section{Acknowledgements}
We would like to thank Karen Abbott for helpful suggestions.

\section{References}
\printbibliography

\end{document}